\documentclass[%
 reprint,
bibnotes,
 amsmath,amssymb,
 aps,
prx,
floatfix,
]{revtex4-2}
\usepackage[export]{adjustbox}
\usepackage{graphicx}
\usepackage{dcolumn}
\usepackage{bm,booktabs,hyperref,textcomp,gensymb,tikz,placeins}
\usepackage{siunitx}
\usepackage{bm}

\usepackage{mathtools}
\DeclarePairedDelimiterX{\infdivx}[2]{(}{)}{
	#1\|#2
}

\newcommand{\rateconstsymbol}{r}
\newcommand{\eqrateconstsymbol}{\kappa}
\newcommand{\br}{\bm{\rateconstsymbol}}
\newcommand{\bk}{\bm{\eqrateconstsymbol}}
\newcommand{\infdiv}{D\infdivx}
\newcommand{\DKLO}{\infdiv{0;\bk'}{\bk}}
\newcommand{\DKLT}{\infdiv{\Theta;\bk'}{\bk}}
\newcommand{\DKLt}{\infdiv{\bm{\theta};\bk'}{\bk}}
\newcommand{\DKLOstar}{D^{*}\infdivx{0;\bk'}{\bk}}
\newcommand{\DKLTstar}{D^{*}\infdivx{\Theta;\bk'}{\bk}}

\newcommand\encircle[1]{%
  \tikz[baseline=(X.base)] 
    \node (X) [draw, shape=circle, inner sep=-1] {\strut #1};}
\newcommand{\circlesymbol}[1]{\protect \encircle{\footnotesize \sffamily #1} \protect}
\newcommand{\Ptot}{\mathcal{P}_\text{comp}}

\newcommand{\Ptrans}{\mathcal{P}_\text{trans}}
\newcommand{\Pss}{\mathcal{P}_\text{ss}}

\begin{document}

\preprint{APS/123-QED}

\title{The information gain limit of biomolecular computation}

\author{Easun Arunachalam}
\affiliation{
 Department of Molecular and Cellular Biology, Harvard University, Cambridge, MA, USA\\
 Green Center for Systems Biology and Lyda Hill Department of Bioinformatics, University of Texas Southwestern Medical Center, Dallas, TX, USA
}
\author{Milo M. Lin}
 \email{Milo.Lin@UTSouthwestern.edu}
\affiliation{
    Green Center for Systems Biology, Lyda Hill Department of Bioinformatics, Department of Biophysics, and Center for Alzheimer's and Neurodegenerative Diseases, University of Texas Southwestern Medical Center, Dallas, TX, USA
}

\date{\today}

\begin{abstract}
    \noindent Biomolecules stochastically occupy different possible configurations with probabilities given by non-equilibrium steady-state distributions. These distributions are determined by the transition rate constants between different configurations. Changing these biochemical parameters (inputs) alters the resulting distributions (outputs), and thus constitutes a form of computation. The information-theoretic advantage of performing computations using non-equilibrium distributions, which require a thermodynamic driving force and thus continual energy expenditure to maintain, is unclear. Here we show how much driving can change probability distributions beyond what is possible at equilibrium. First, we establish a tight limit on how much the driving force can change the probability of observing any configuration of an arbitrary molecular system. We then derive a concise expression relating the driving force to the maximum information gain -- the change in the full probability distribution over configurations -- in any computation, showing how small input changes can exponentially alter outputs. Finally, we numerically show that synthetic systems and Ras signaling can closely approach this bound, illustrating the necessity of energy expenditure to enable the computational capabilities observed in nature.
\end{abstract}

\maketitle

\section{\label{sec:level1}Introduction}

    Biomolecules can exist in different states with distinct functionalities: nucleic acids and proteins can adopt different conformations, they can bind to each other or to ligands to form complexes, and they can be phosphorylated and acetylated. Switching between these different possible states, often in response to input signals from the environment, constitutes a general type of biomolecular computation \cite{bray1995protein,takaki2022information} that is distinct from classical notions of computation.
    
    Classical models of computation involve modifying a single copy of a rewritable memory that remains unchanged until subsequent operations are performed. In contrast, biomolecular computation typically depends on the ensemble- and time-averaged configurations of many copies of a molecule, and consists of transitions between different steady-state probability distributions of molecular configurations. Individual computational steps consist of a change in molecular rate constants (inputs) from $\br \to \br'$ which leads to an updated probability distribution (outputs; Fig. \ref{fig:fig1}A). In cases such as protein folding \cite{sali1994does, tehver2008kinetic}, modulation of enzyme activity \cite{otten2020directed}, or signaling \cite{latorraca2017gpcr,simanshu2017ras}, there is often a single functionally relevant ``on'' state whose probability is updated by the computation. In other cases, such as self-assembly \cite{brugues2012nucleation,morrow2015ph}, biologically relevant information is encoded by the entire probability distribution over all states.

    The total power consumption associated with biomolecular computation (i.e. the energetic cost per unit time), which we denote by $ \Ptot $, can formally be decomposed into two components: that associated with the transient dissipation when transitioning from one steady-state distribution to another in response to a change in inputs ($ \Ptrans $), and that associated with maintaining the steady-state distribution after the transient dissipation ($ \Pss $):
    \begin{equation}
        \Ptot = \Ptrans + \Pss
        \label{eq:energetic_cost_decomposition}
    \end{equation} 
    (Note that we define the cost of computation to include only the power consumption associated with changing or maintaining the system; the power consumption required to change the inputs is considered to be a separate cost.)
    Although Landauer acknowledged the general importance of $\Pss$ in his seminal work \cite{landauer}, the dynamical stability of magnetic storage devices means that steady-state dissipation is typically negligible compared to transient dissipation in human-made computers ($ \Pss \ll \Ptrans $). In such cases, the minimal energetic cost of bit erasure in a logically irreversible computational step, which is the time integral of $\Ptrans$, is $kT\ln2$ \cite{landauer,toffoli1982conservative, bennett1982thermodynamics,bennett2003notes} where $ kT $ is the product of the Boltzmann constant and the temperature.
    
    In contrast, $\Pss$ is a major cost associated with computation in living systems, which transition between non-equilibrium steady states that require energy to maintain. This requires cells to harness thermodynamic forces, often in the form of an excess concentration of an energy currency such as adenosine or guanosine triphosphate (ATP or GTP) \cite{mugnai2020theoretical}. In the cell, these forces are typically kept constant while the system changes its other parameters (which we refer to as \textit{passive parameters}) from $\bk \to \bk'$ in order to convert between non-equilibrium steady states. Such input changes can be accomplished, for example, via changes in the concentrations of ligands or cofactors, or reversible covalent modifications such as phosphorylation. Whereas previous results mostly focused on $\Ptrans$, in this work we address how the thermodynamic forces that generate $ \Pss $ constrain information processing in biomolecular computation. We derive a universal relation showing how information gain in any molecular computation describable by a master equation is limited by the thermodynamic force that maintains the non-equilibrium steady states.

    \subsection{Computation using equilibrium versus non-equilibrium steady-states}
    
    \begin{figure*}[!htpb]
        \includegraphics[width=0.65\textwidth]{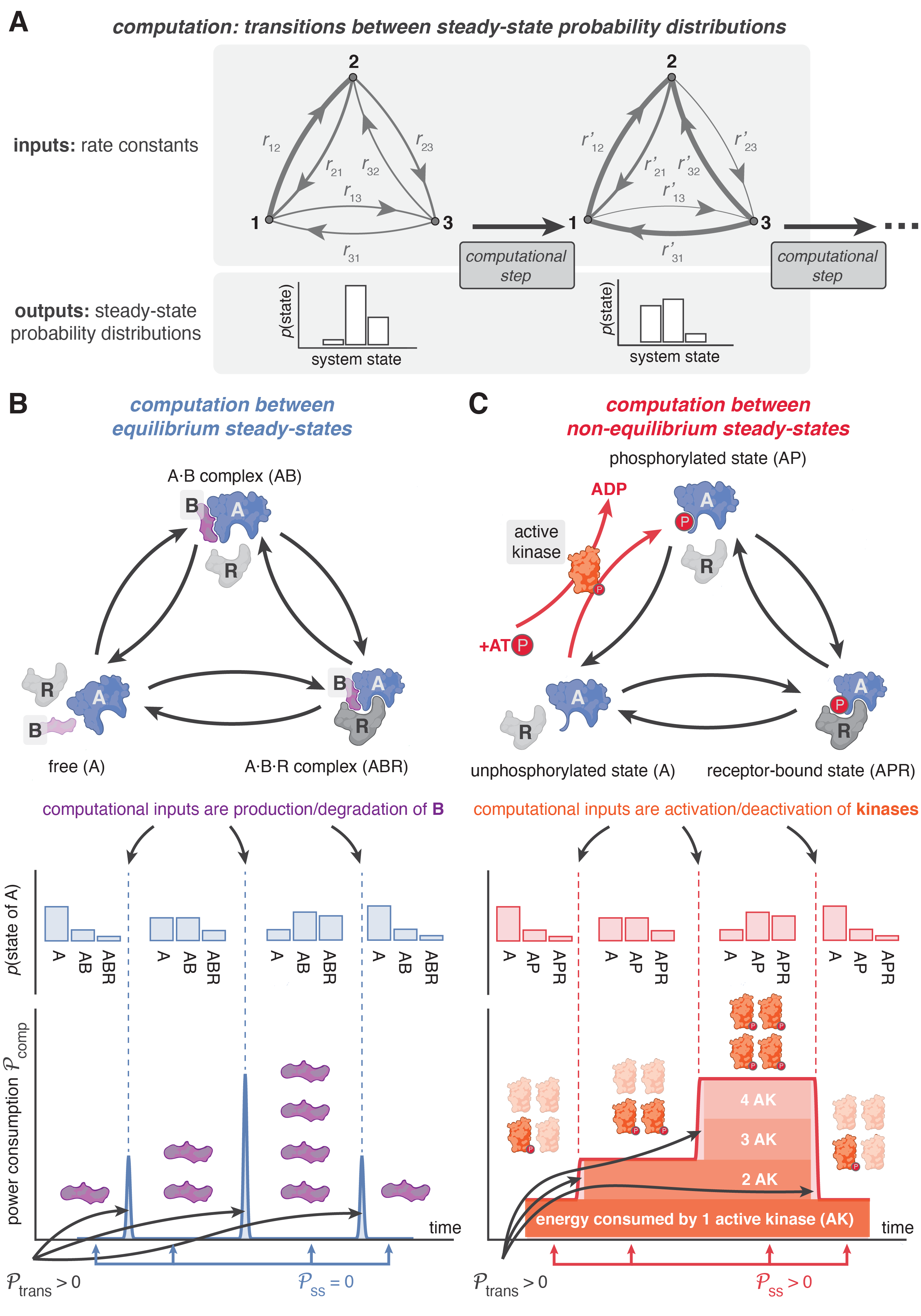}
        \caption{\label{fig:fig1} \textbf{Steady-state computation in molecular biology.}
        (A) Changing the steady-state probability distribution of system states constitutes a computational step.
        (B) Protein-receptor binding is an example of computation between equilibrium steady-states.
        (C) Protein phosphorylation-gated receptor binding is an example of computation between non-equilibrium steady-states.
        }
    \end{figure*}
    
    To understand whether computation between non-equilibrium steady states, which are energetically costly to maintain, confers quantifiable advantages over computation between equilibrium steady-states, we first illustrate the difference between these two types of computation.
    Protein complex formation is a prototypical example of computation between equilibrium steady-states (Fig. \ref{fig:fig1}B). In a three-state model of this process, protein \textit{A} can reversibly bind protein \textit{B}, the \textit{A}$ \cdot $\textit{B} complex can reversibly bind a receptor \textit{R}, and the \textit{A}$ \cdot $\textit{B}$ \cdot $\textit{R} complex (i.e. the ``on'' state) can reversibly dissociate.
    The relative populations of the different states of $ A $ can be changed by increasing the number of \textit{B} proteins or altering them (e.g. via covalent modifications or ligand binding) to increase their binding affinity to \textit{A}, which will increase the probability of finding \textit{A} in the ``on'' state. Operation of this system requires energy expenditure only during the computational steps, when the system is transitioning to new steady states due to changing inputs ($ \Ptrans > 0 $). Because the steady states are at equilibrium, the forward flux equals the backwards flux for every transition. Consequently, energy is not required to maintain the steady state distributions when inputs are not changing ($ \Pss = 0 $) (Fig. \ref{fig:fig1}B).
 
	Phosphorylation-gated binding is a prototypical example of computation between non-equilibrium steady-states (Fig. \ref{fig:fig1}C). An active kinase can consume energy to phosphorylate protein \textit{A}, and a phosphatase can dephosphorylate \textit{A}; phosphorylated \textit{A} can reversibly bind a receptor \textit{R}; and reversible dephosphorylation not mediated by an enzyme can lead to receptor unbinding. In contrast to computation between equilibrium steady-states, the power consumption of this system has two distinct components: the power dissipated to reach new steady states after a  change in the inputs ($ \Ptrans > 0 $) and the power required to maintain the non-equilibrium steady-state, characterized by net clockwise probability flux that breaks detailed balance, even when inputs are not changing ($ \Pss > 0 $) (Fig. \ref{fig:fig1}C). Our goal is to quantify how the sensitivity of the output distribution to changing inputs depends on whether the steady states are at equilibrium (e.g. Fig. \ref{fig:fig1}B) or not (e.g. Fig. \ref{fig:fig1}C).

    \subsection{Quantifying computation between steady-states}
    
    We consider discrete-state continuous-time Markov models describing mass-action kinetics of biomolecules coupled to a bath at temperature $ T $. Systems at equilibrium are parameterized by \textit{passive rate constants} $ \eqrateconstsymbol_{ij} $ associated with transitions between states $i$ and $j$ . The probability of the system occupying state $ i $ at time $t$, denoted by $ {p}_i(t) $, evolves according to
    \begin{align}
        \frac{d{p}_i(t)}{dt} = \sum_{j\neq i}\big(\eqrateconstsymbol_{ji} {p}_j(t) - \eqrateconstsymbol_{ij} {p}_i(t)\big).
    \end{align}
    The steady-state value of $p_i(t) $ is denoted by $ p_i $. 
    
    At equilibrium, the net probability current between $ i $ and $j $, denoted by $ I_{ij} = \eqrateconstsymbol_{ij} p_i - \eqrateconstsymbol_{ji} p_j $, is zero for all $ ij $ (the condition of detailed balance). Non-equilibrium systems can be constructed by starting from a reference equilibrium system and changing some of the forward rate constants from the passive rate constant $\eqrateconstsymbol_{ij}$ to the driven rate constant $r_{ij}$, which in general will break detailed balance. We shall frequently compare a non-equilibrium system to its reference equilibrium system.
    \par
    We define the \textit{applied driving force} $ \theta_{ij}$ in terms of the driven and reference passive rate constants between states $i$ and $j$:
    \begin{equation}
        \theta_{ij} = k T \ln \left( \dfrac{r_{ij}}{\eqrateconstsymbol_{ij}} \right)
        \label{eq:def_theta}
    \end{equation}
    Note that $ \theta_{ij}$ should be thought of as energy per molecular transition rather than energy per length and so has units of energy. The applied driving forces $ \theta_{ij}$ are also related to chemical potential differences: the sum of the net applied driving force along any loop is equal to the net chemical potential difference along that loop, which is also called the cycle affinity \cite{schnakenberg1976network, seifert2012} (Appendix \ref{sec:theta_mu_rel}). Applied driving forces are directly related to the steady-state power consumption $ \Pss $ of the system \cite{lin2024general}\footnote{Scaling all rate constants in the system by a proportionality factor does not change the steady-state probability distribution but does change the power consumption; thus, the relevant, constraining quantity is the driving force $ \theta $.}:
    \begin{equation}
        \Pss = \sum_{mn} \theta_{mn} I_{mn}.
        \label{eq:pss_definition}
    \end{equation}
    where $I_{mn}$ are the net probability currents induced by the driving forces, and the sum is performed over each pair of connected states $ m $ and $n $. Note that, in general, dissipation occurs at all transitions even though Eq. \ref{eq:pss_definition} shows that the total dissipation can be calculated by considering only the driven transitions, for which $\theta_{mn} \neq 0$. 

    In contrast to the steady-state dissipation rate $\Pss$ or the chemical potential difference $\Delta \mu$ between two states, which are generally emergent properties of the system, the applied driving force $\theta$ between two states is an explicit function of the rate constants between those states. Therefore, $\theta$ serves as a directly controllable design parameter. For example, in biological systems, $ e^{\theta/kT} $ is often the fold-increase of ATP concentration over its concentration at equilibrium. This driving leads to net entropy production in the thermal bath and requires constant external supply of energy to regenerate ATP and maintain the steady state.

    We denote by $ \bm{\theta} $ the set of all applied driving forces $ \theta_{ij} $ for all $ ij $, and by $ \bk $ the set of all passive parameters $ \eqrateconstsymbol_{ij} $ defining the reference equilibrium system for all $ ij $. As can be seen in Eq. \ref{eq:def_theta}, each rate constant of the system $ \rateconstsymbol_{ij}$ is uniquely determined by $ \eqrateconstsymbol_{ij} $ and $\theta_{ij}$; the steady-state probability of state $ i $ is thus represented by $ p_i( \bm{\theta}, \bk) $. In most contexts, biomolecular computation corresponds to changing the passive parameters $ \bk $, while keeping the non-equilibrium driving forces $ \bm{\theta}$ (e.g. ATP/ADP ratio) fixed. We formally define a computational step as the transition $ p_i( \bm{\theta}, \bk) \to p_i( \bm{\theta}, \bk')$ in response to changing the passive input parameters from $ \bk \to \bk' $. We seek to determine, in the most general case, which input-output relationships are possible in such a computational step. This can be rigorously formulated in the language of information theory as the information gain \cite{murphy2022probabilistic} (also known as the Kullback-Leibler divergence, or relative entropy) $ D $ between the steady-state probability distribution of the post-computation system and the probability distribution of the pre-computation system:
    \begin{equation}
        \DKLt = \sum_{i=1}^N p_i(\bm{\theta},\bk')\ln{\left[\dfrac{p_i(\bm{\theta},\bk')}{p_i(\bm{\theta},\bk)} \right]}.
    \end{equation}
    The information gain is the expected excess surprise (as measured in bits per random sampling of the configuration of the system) from continuing to use $p_i(\bm{\theta},\bk)$ as the probability distribution even though the probability distribution should be updated to $p_i(\bm{\theta},\bk')$ due to the change of input conditions $ \bk \rightarrow \bk' $. The information gain is not the change in entropy of the distribution, which can be positive or negative. Rather, it is a measure of the error in using the pre-computation distribution as a model of the post-computation distribution, which is always non-negative.

    We find through numerical experimentation that the information gain is sensitive to the applied driving force within a limited range, beyond which it saturates (Appendix \ref{sec:circuit}). Below, we derive mathematical relations that precisely establish how the probability amplification of any state $ p_i (\bm{\theta},\bk) / p_i (\bm{0},\bk)$, as well as the information gain over all states $ \DKLt $, are limited by the driving forces $ \bm{\theta} $ for arbitrary computational steps $ \bk \rightarrow \bk'$ in any system. Information gain and suppression of noise arising from $ \bk \to \bk' $ parameter fluctuations are shown to be equally constrained by the thermodynamic force; they are fundamentally symmetric sides of the same coin.

    \section{Results}
    
	\subsection{Applied driving forces bound amplification and suppression of individual states}
    
    The components of information gain are the probabilities of individual states. Thus, to obtain our main result, we first prove bounds on how the probabilities of individual states can change under non-equilibrium driving.

    The steady-state probability of any state can be expressed in terms of rate constants present in distinct directed spanning trees of the network of states. The relevant object then becomes non-self-intersecting paths between two states; the presence of many driven transitions in a single path enables extreme probability amplification and suppression. We define the total applied driving force $ \Theta $ to be the maximum sum of applied driving forces along any non-self-intersecting path in the network:
    \begin{equation}
        \Theta \equiv \max_{ \text{n.s.i. path } u } \left( \sum_{ \text{edge } mn \: \in \: u} \left( \theta_{mn} - \theta_{nm}\right) \right)
        \label{eq:Theta_definition}
    \end{equation}
    In Appendices \ref{sec:amp_lim_proof}-\ref{sec:amp_lim_connection}, we use Kirchhoff's matrix-tree theorem \cite{schnakenberg1976network} to show our first main result: the steady-state probability of any state $ i $ in the presence and absence of the applied driving force ($ p_i(\Theta,\bk) $ and $ p_i(0,\bk) $, respectively) is constrained by:
    \begin{align}
        p_i(0,\bk) e^{-\Theta / k T} \leq p_i(\Theta,\bk) \leq p_i(0,\bk) e^{\Theta / k T} .
        \label{eq:amplimit_maintext}
    \end{align}
    Hence, $ \Theta $ limits the maximal amplification or suppression of any state. This bound is general, and applies to the steady-state behavior of all master equation systems regardless of the complexity of the network. This result differs from previous work \cite{blowtorch2013Maes, gunawardena2022blowtorch, liang2024thermodynamic} bounding the relative probabilities between two different states of the same system. Other recent work produced a bound that implies Eq. ~\ref{eq:amplimit_maintext} in the case of a single driven transition \cite{cetiner2023universal}. In contrast, Eq. ~\ref{eq:amplimit_maintext} bounds the change in the probability of the same state if the system is altered by an arbitrary set of thermodynamic driving forces, even across multiple driven transitions (Appendices \ref{sec:amp_lim_proof}-\ref{sec:amp_lim_connection}).

    \begin{figure*}[htpb]
        \includegraphics[width=0.8\textwidth]{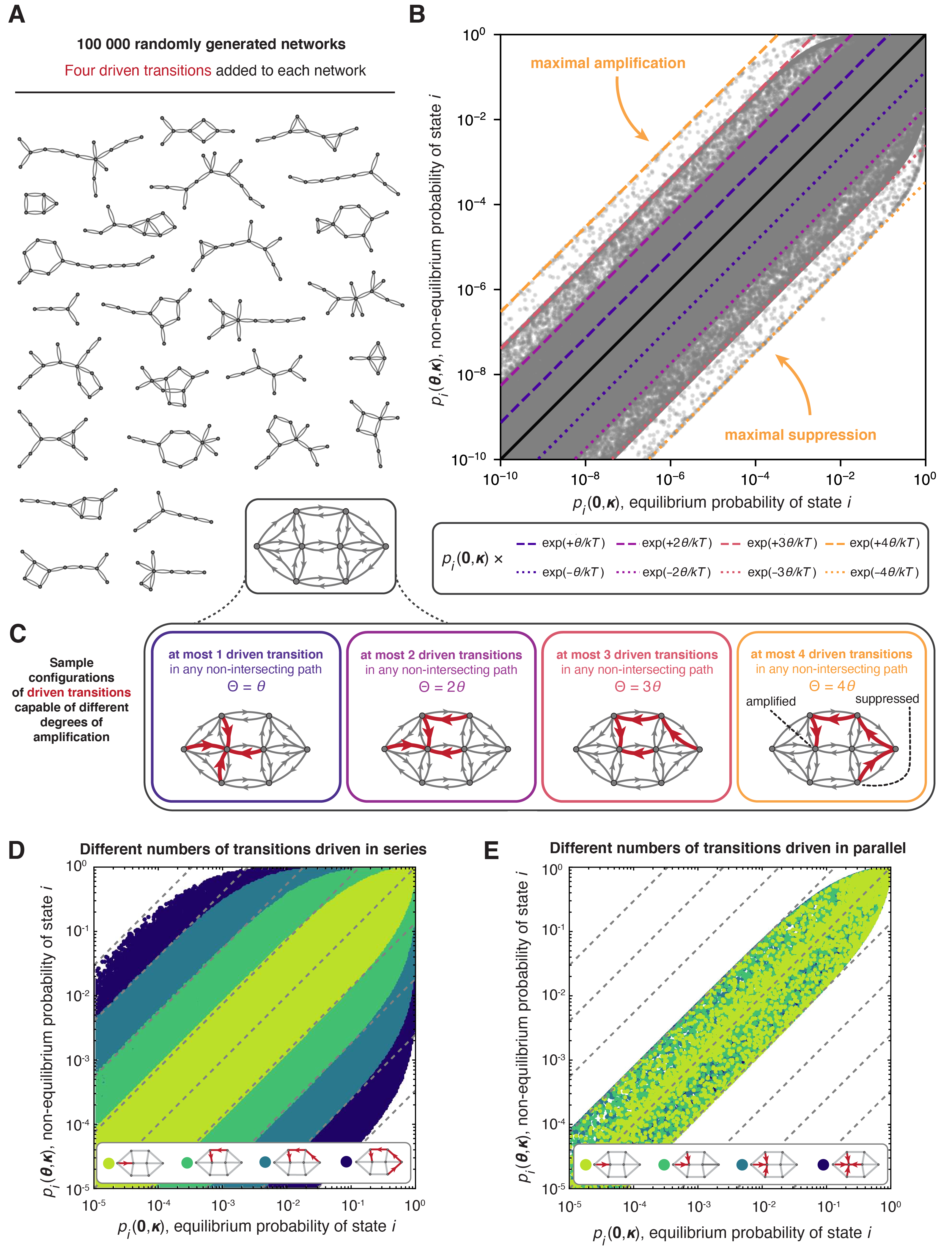}
        \caption{\label{fig:fig2} \textbf{Total applied driving force tightly bounds probability amplification and suppression in arbitrary networks.} (A) Examples of network topologies whose probability amplification and suppression properties were studied. Four driven transitions were added in different locations in each network, and the equilibrium and non-equilibrium probabilities of each state were calculated. (B) Equilibrium and non-equilibrium probabilities of individual states in each network. Each dot represents a single state. (C) Different arrangements of the four driven transitions lead to differences in the maximum number of driven transitions along any non-intersecting path (i.e. different total applied driving forces), and hence different bounds on probability amplification and suppression. (D) Equilibrium and non-equilibrium probabilities of individual states in networks with different numbers of transitions driven in series. (E) Equilibrium and non-equilibrium probabilities of individual states in networks with different numbers of transitions driven in parallel.}
	\end{figure*}

    To illustrate the tightness of this bound, we generated a range of state-space networks with diverse topologies and passive parameters $ \bk $ (Fig. ~\ref{fig:fig2}A). In each system, we randomly selected four transitions along which to apply a non-equilibrium driving force ($ \theta = 2 \: k T $) and compared the equilibrium probability $ p_i(\bm{0},\bk) $ and non-equilibrium probability $ p_i(\bm{\theta},\bk) $ of each state $ i $. We observe that $ p_i(\bm{\theta},\bk) $ cannot exceed $ p_i(\bm{0},\bk) e^{4 \theta / k T} $, and cannot be lower than $ p_i(\bm{0},\bk) e^{-4 \theta / k T} $. Moreover, there are many states for which $ p_i(\bm{\theta},\bk) $ lies along discrete sub-bounds given by $ p_i(\bm{0},\bk) e^{\pm \theta / k T} $, $ p_i(\bm{0},\bk) e^{\pm 2 \theta / k T} $, and $ p_i(\bm{0},\bk) e^{\pm 3 \theta / k T} $. Our result (Eq. \ref{eq:amplimit_maintext}) explains the common origin of the observed overall bound and sub-bounds as follows.
    In the example set of systems containing four driven transitions analyzed in Fig. ~\ref{fig:fig2}, there may exist arrangements in which all four driven transitions point in the same direction along a single non-intersecting path. In this case, $ \Theta = 4 \theta $, and some states may be maximally suppressed or amplified. However, for other arrangements, there may not exist any non-intersecting paths containing all of the driven transitions. If at most one, two or three driven transitions are part of a non-intersecting path, we have $ \Theta = \theta $, $ 2 \theta $ or $ 3 \theta $, corresponding to the discrete sub-bounds observed in the simulations (Fig. \ref{fig:fig2}C). Thus, even though there may exist four driven transitions in the overall network, no state probability can be suppressed or amplified to the extent possible when $ \Theta = 4 \theta $ (Fig. ~\ref{fig:fig2}C).

    This result also explains why the maximum amplification or suppression depends on whether driven transitions are arranged in a series or parallel configuration. When arranged in series, a single non-intersecting path can contain all transitions, and hence amplification depends on the number of driven transitions  (Fig. ~\ref{fig:fig2}D). In contrast, when arranged in parallel (e.g. transitions from several different states are directed towards the same state), no path can contain more than one driven transition pointing the same direction along the path. Hence, the maximum amplification is determined by the largest driving force among the different transitions, regardless of the total number of driven transitions (Fig. ~\ref{fig:fig2}E). Overall, the bound for each value of $ \Theta $ (Eq. ~\ref{eq:amplimit_maintext}) is tight.

    \subsection{Maximum signal amplification and suppression limit}
    Eq. ~\ref{eq:amplimit_maintext} bounds the change in probability of any state of a system if the driving force $\Theta$ is applied. We next used this relation to derive bounds relevant for biomolecular computation, in which the passive parameters $\bk$ are changed while $\Theta$ is held constant.

    The maximum amplification or suppression of the probability of any state $i$ after a computational step $ \bk \to \bk' $, relative to the amplification of the same computational step at equilibrium, is given by:
    \begin{equation}
        \Bigg|\ln{\left(\dfrac{p_i(\Theta,\bk)}{p_i(\Theta,\bk')}\right)} - \ln{\left(\dfrac{p_i(0,\bk)}{p_i(0,\bk')}\right)} \Bigg| \leq \dfrac{2\Theta}{k T}.
        \label{eq:sigratio_bound}
    \end{equation}
    This amplification/suppression relation applies to the change in probability of any single biologically relevant state $i$ of a system (e.g. a signaling ``on'' state). An input event, such as phosphorylation, changes the passive parameters of the signaling protein from $ \bk \to \bk' $, thereby altering the probability of the signaling state, which conveys the signal to downstream components of the signaling pathway. We consider the specific case of the Ras signaling protein below.
    
    \subsection{Maximum and minimum information gain limit}

    In other cases such as self-assembly, the overall probability distribution over all system states is biologically meaningful. Here, rather than the log-fold-change of an individual state, the relevant measure of computation is the information gain, which is the log-fold-change averaged over all states. We used Eq. ~\ref{eq:amplimit_maintext} to derive bounds on non-equilibrium information gain ($\infdiv{\Theta;\bk}{\bk'}$) as a function of the equilibrium information gain ($\infdiv{0;\bk}{\bk'}$) and the total applied driving force $ \Theta $ (Appendix \ref{sec:infogain_limit}):
    \begin{align}
        \begin{split}
            \text{upper bound: }& \DKLT \leq \dfrac{2\Theta}{k T} + \DKLO \mathrm{e}^{\Theta / k T} \\
            \text{lower bound: }& \DKLO \leq \dfrac{2\Theta}{k T} + \DKLT \mathrm{e}^{\Theta / k T}
        \end{split}
        \label{eq:infogainbound}
    \end{align}
    The bound on non-equilibrium information gain $ \DKLT $ has two components. The first term, which is independent of the equilibrium information gain and scales linearly with the thermodynamic force, is relevant to computational steps such as the addition of a catalyst that acts on the system. Such steps lead to zero information gain at equilibrium but non-zero information gain in the presence of the thermodynamic force. Because catalyst molecules are present at sub-stoichiometric levels, multiple types of catalysts can simultaneously act on the system to perform complex input-output computations without overcrowding. 
    For example, the concentrations of some kinases are more than three orders of magnitude lower than their target substrates \cite{martins2013ultrasensitivity}. The second term is proportional to the equilibrium information gain and scales exponentially with the thermodynamic force. However, this exponential amplification can only act on information gain already present when $ \Theta = 0 $ (computation between equilibrium steady-states) and must therefore be implemented via stoichiometric modifications.
    
    \begin{figure}[htpb]
        \includegraphics[width=0.45\textwidth]{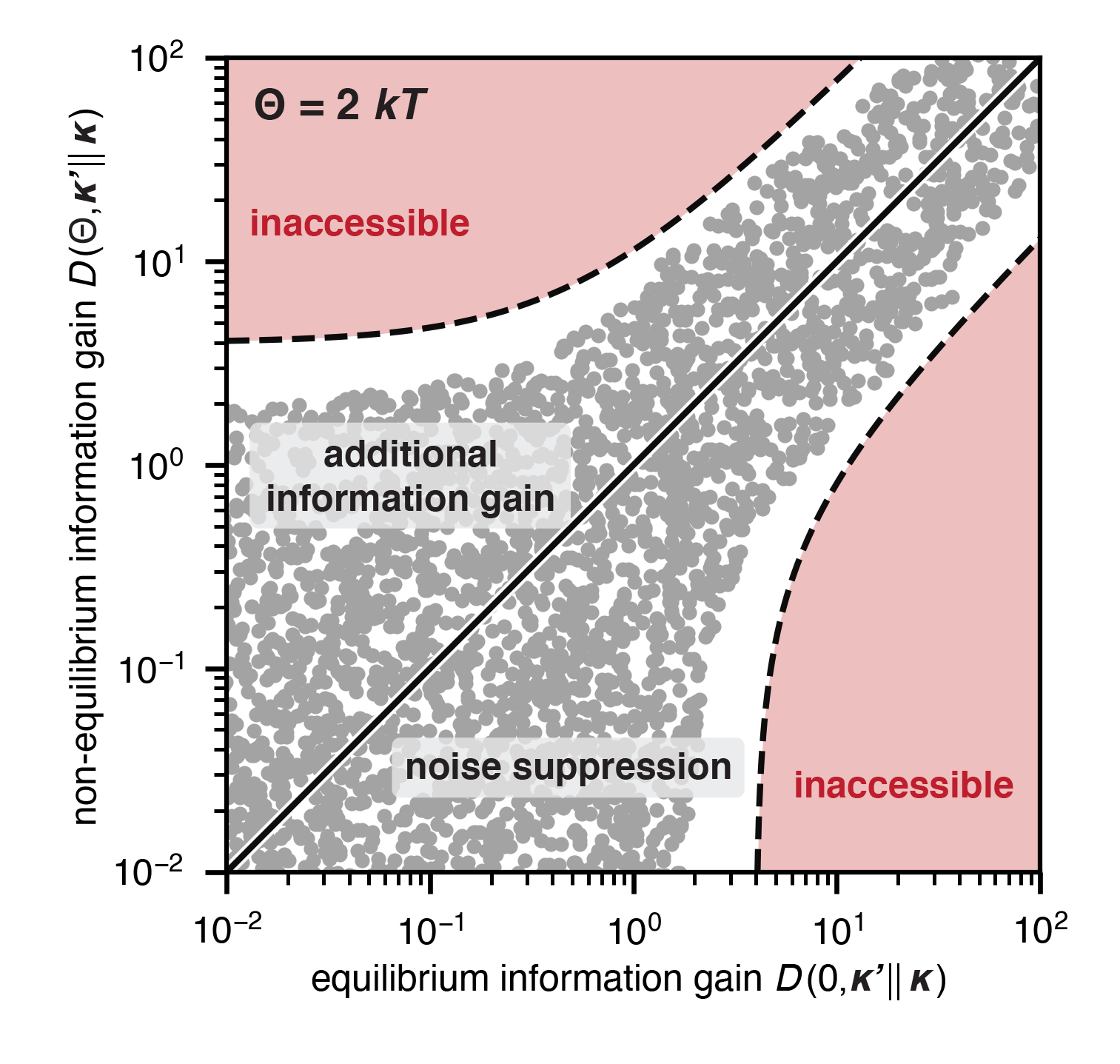}
		\caption{\label{fig:fig3a} \textbf{\small Accessible regimes of non-equilibrium information gain.}
        Each gray point represents a pair of systems (with pre-computation parameter set $ \bk $ and post-computation parameter set $ \bk' $). Bounds are indicated by dashed lines. The region above the diagonal corresponds to additional information gain (information gain due to the $\bk \to \bk'$ transition is greater when the systems are driven) and the region below the diagonal corresponds to noise suppression (driving reduces the information gain associated with the $\bk \to \bk'$ transition). The red shaded regions indicate inaccessible regions.
        }
	\end{figure}

    Some systems can consume energy to generate larger changes in their state probability distributions than would be possible through modification of passive parameters $ \bk $ alone, resulting in $ \DKLT > \DKLO $. Such systems are capable of \textit{additional information gain} that is enabled by non-equilibrium driving. Other systems consume energy to avoid changes in their state probability distributions that would occur when the underlying passive parameters were modified from $ \bk \to \bk' $, i.e. $ \DKLT < \DKLO $. In such cases, non-equilibrium driving enables \textit{noise suppression}, making systems robust against fluctuations in $ \bk $. Eq. ~\ref{eq:infogainbound} provides upper and lower bounds on additional information gain and noise suppression, respectively.

    \begin{figure*}[!htpb]
        \includegraphics[width=0.85\textwidth]{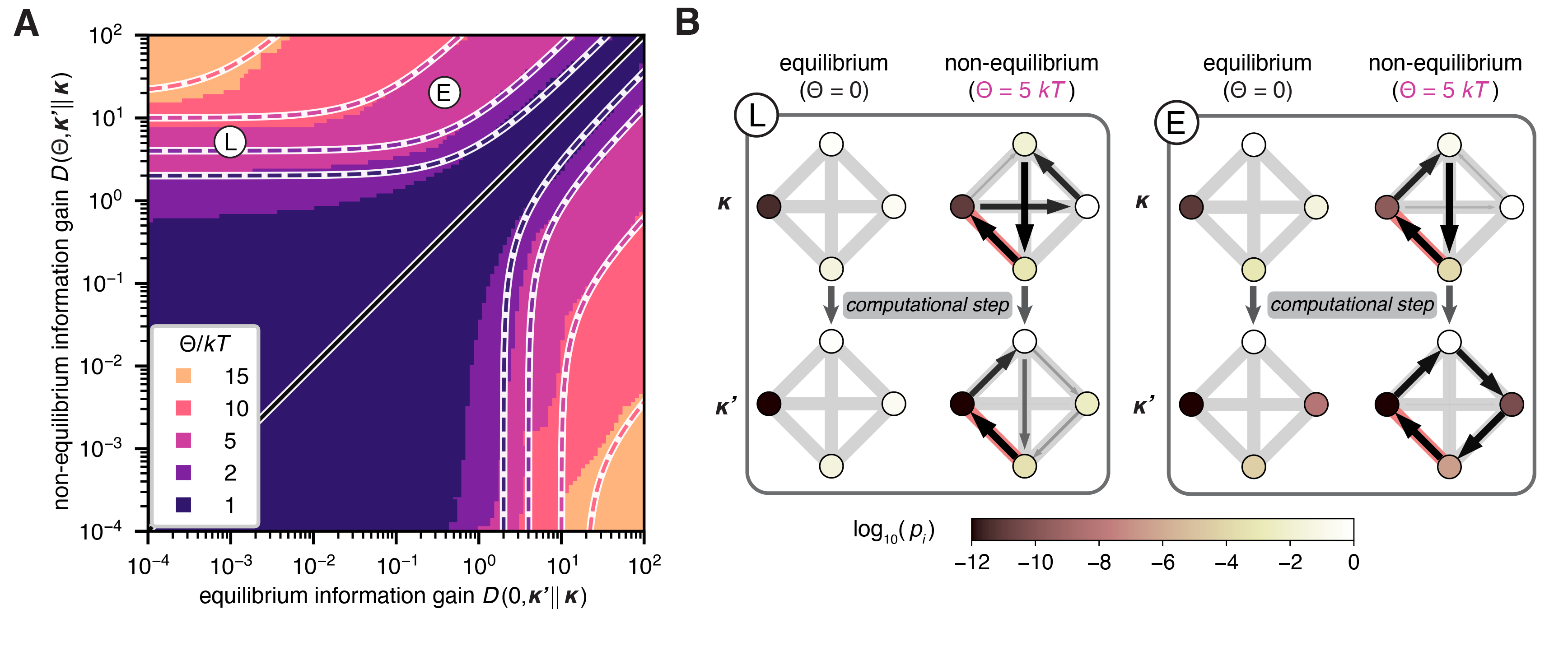}
		\caption{\label{fig:fig3b} \textbf{\small Numerical optimization illustrates the tightness of the information gain bound both near and far from equilibrium.}
        (A) Achievable values of equilibrium and non-equilibrium information gain are shown for different values of the total applied driving force $ \Theta $. Each filled point corresponds to a pair of passive parameters, and dashed lines denote the corresponding bound for each value of $ \Theta $. The locations of representative systems shown in (C) are indicated by \circlesymbol{L} and \circlesymbol{E}.
        (B) Representative pairs of four-state completely connected systems with an applied driving force of $ \Theta = 5 \: k _B T $ in the linear \circlesymbol{L} and exponential \circlesymbol{E} regimes.
        }
	\end{figure*}

    We tested the tightness of the bound in Eq. ~\ref{eq:infogainbound} for a total applied driving force of $\Theta = 2 kT$ by constructing a series of state-space networks with different topologies. For each network, we considered two pairs of passive parameters $ \bk $ and $ \bk' $, which we optimized to maximize or minimize $ \DKLT $ given a particular value of the equilibrium information gain $ \DKLO $ (shown in gray in Fig. ~\ref{fig:fig3a}) (Appendix \ref{sec:mcopt}). While our simulations did not approach the bound arbitrarily tightly, optimization brings them close to the upper and lower bounds.

    \subsection{The information gain bound is relevant far from equilibrium}

    Having verified the tightness of the bound for a modest value of the applied driving force, we sought to test whether the bound was also relevant far from equilibrium. Focusing on a completely connected four-state network, we optimized $ \bk $ and $ \bk' $ for a range of applied driving forces up to $ 15 \: k T $, comparable to the driving force supplied by the hydrolysis of ATP or GTP \cite{nelson2008lehninger}. We found that the bound was similarly tight even for systems driven far from equilibrium (Fig. ~\ref{fig:fig3b}A).

    We examined representative sets of passive parameters $\bk$ for the four-state complete network that generated information gain values in the regions where either the linear or exponential terms in the bound dominate for $\Theta = 5  \: k T $ (Fig. \ref{fig:fig3b}B). In the linear region, the steady-state probability distributions for the $ \bk $ and  $ \bk' $ parameter sets are nearly identical for the equilibrium case, and are different in the non-equilibrium case. The probability currents also change qualitatively from $ \bk $ to  $ \bk' $. In the exponential region, the probability distribution differs slightly between $ \bk $ and $ \bk' $ in the equilibrium case, and changes drastically in the non-equilibrium case; again, the pattern of currents changes. Taken together, these data indicate that the changes in passive parameters needed to closely approach the bound are non-trivial, leading to qualitative variation in probability and current distributions in different information gain regimes.

    \subsection{The signal amplification bound constrains biological phenomena}
    
\begin{figure*}[!htpb]
        \includegraphics[width=0.8\textwidth]{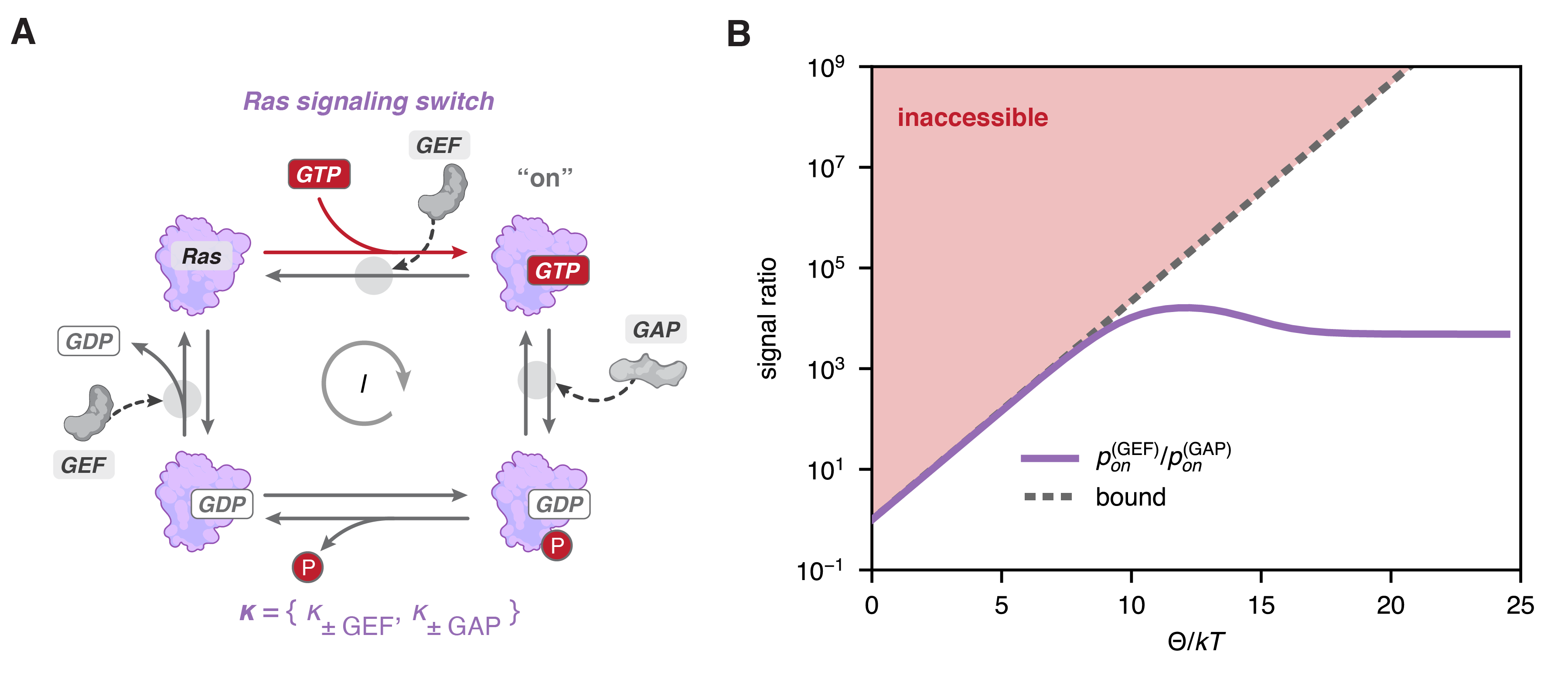}
        \caption{\label{fig:fig4} \small \textbf{Comparison of Ras signaling to signal amplification bound.} (A) Structure of the Ras signaling network. Cofactors are omitted from reverse reactions for clarity. GAP and GEF each alter rate constants of specific processes; their presence or absence are the control parameters of the system. (B) The ratio of probability of the ``on'' state in the presence of GEF versus the presence of GAP, or the ``signal ratio'', as a function of applied driving force.}
	\end{figure*}

    Having shown that synthetic systems can closely approach the bounds on signal amplification and information gain, we analyzed a biological example with reported rate constants: the Ras signaling switch, which transduces signals controlling cell growth and differentiation. Ras is a protein whose work cycle can be modeled as a four-state system with one driven transition, consisting of GTP binding and hydrolysis, and GDP and phosphate unbinding (Fig. ~\ref{fig:fig4}A) \cite{neal1988kinetic}. The presence of a guanine nucleotide exchange factor (GEF) affects the rate of GTP and GDP (un)binding, and the presence of a GTPase activating protein (GAP) affects the rate of GTP hydrolysis. Thus, the presence or absence of GEF and GAP act as tunable inputs that change the passive parameters of the system, and hence the probabilities of the states (Fig. \ref{fig:supp_fig_ras}).

    The probability of the GTP-bound ``on'' state, in which Ras can activate downstream machinery, is the biologically meaningful quantity. Hence, the relevant bound is on the amplification of this state. For the specific case of Ras, in which the driven transition points directly to the ``on'' state, Eq. ~\ref{eq:sigratio_bound} reduces to a tighter bound: $ \big|\ln{\left(\frac{p_i(\Theta,\bk)}{p_i(\Theta,\bk')}\right)} - \ln{\left(\frac{p_i(0,\bk)}{p_i(0,\bk')}\right)} \big| \leq \frac{\Theta}{k T} $ (Appendix \ref{sec:ras}). Using fixed rate constant values which were previously measured (Table \ref{tab:Rasparams}), we asked to what extent GAP or GEF can alter the probability of the ``on'' state, quantified by the signal ratio of the ``on'' state probability under each of the input conditions: $p^{(\mathrm{GEF})}_{on}/p^{(\mathrm{GAP})}_{on}$. Fig. ~\ref{fig:fig4}B shows that the signal ratio is tightly bounded by the theoretical maximum for low and medium values of the applied driving force, and reaches a plateau at higher values.
    We hypothesize that driving beyond this level yields diminishing returns because further increases in the signal ratio exceed the dynamic range of the output signal.
    In this regime, the plateau in the response confers robustness to fluctuations in GTP and GDP levels under physiological conditions ($ \Theta \approx 20  \: kT $).
    
    \section{Discussion}
    
    Previous work has highlighted how non-equilibrium driving affects information processing in biological systems. For example, it has been shown that energy expenditure generates the sharp morphogen profiles necessary for eukaryotic development which are inaccessible at equilibrium \cite{estrada2016information} and enables error reduction in copying processes \cite{hopfield1974kinetic,song2020thermodynamic}.
    Considerable theoretical work has focused on understanding how energy expenditure constrains the fluctuations of a system \cite{crooks1999entropy,barato2015thermodynamic,gingrich2016dissipation,paulsson2010fundamental,ohga2023thermodynamic}.
    
    Yet, fluctuations are distinct from the transitions between steady-states that constitute computation. The few studies on the thermodynamic constraints on computation have focused on transient behaviors or perturbative changes \cite{horowitz2017minimum, owen2020universal}.
    However, it remains unclear how applied driving forces constrain changes in probability distributions in response to arbitrary changes in passive system parameters.
    This fundamental question about the thermodynamic limits on biomolecular computation is not addressed by the Landauer energy requirement for irreversible computation \cite{landauer}, which does not apply to the thermodynamic cost of holding probability distributions at fixed non-equilibrium steady-states. 
    
    Recently, a tight bound on information gain in response to driving a single transition was derived \cite{cetiner2023universal}. However, tight bounds on the behavior of multiply driven systems -- which are ubiquitous in biology -- have remained elusive. Moreover, because thermodynamic driving forces such as ATP concentration are typically fixed, living systems alter probability distributions over system states by changing passive system parameters.
    Thus, formulating bounds on information gain in terms of changes in passive parameters is particularly important, and has remained an open problem.
    
    Here we have derived bounds on how much systems can change the probability of individual states and alter their overall information content depending on the strength of the total applied driving force. These bounds apply to the steady-state behavior of all master equation systems regardless of size, complexity, number of driven transitions, or the nature of the computational step. We demonstrate that the probability amplification and suppression bound is tight, and that diverse molecular systems can closely approach the information gain bound, which is relevant for systems both near and far from equilibrium.
    
    Building on pioneering work by Berg and Purcell \cite{berg1977physics}, more recent work \cite{lang2014thermodynamics} has shown that energy consumption limits the sensitivity of molecular sensing. In a similar spirit, we show that the applied thermodynamic force limits the information gain of molecular computation. It is important to note that while information processing in the cell is constrained by energy consumption, in practice it is also limited by noisy integration of signals and sub-sampling of probability distributions \cite{razo2020first}.
    
    We show that a common example of biological signaling can function near the limit of signal amplification derived here. The plateau of the signal ratio and information gain near the biologically relevant values of $ \Theta \approx 15-20 \: kT $ \cite{nelson2008lehninger,zala2017advantage} suggests robustness to natural fluctuations of intracellular GTP and GDP levels, which can vary the driving force by about $ 1 \: kT $ \textit{in vivo} \cite{ahn2017temporal}.
    
    Given the energetic efficiency of computation between equilibrium steady-states, it is perhaps surprising that biological systems rely so heavily on non-equilibrium processes. Our analysis offers a general explanation for this strategy, and examples of systems that can take advantage of it. Future work will reveal the extent to which living systems consistently exploit the additional information gain available when operating far from equilibrium.

\begin{acknowledgments}
    We thank Daniel Busiello, Peter Foster, Todd Gingrich, Jordan Horowitz, Vinothan Manoharan, Daniel Needleman and Yong Hyun Song for valuable feedback. We acknowledge the Texas Advanced Computing Center (TACC) at the University of Texas at Austin for providing computational resources. This work was supported by the Chan Zuckerberg Initiative Theory in Biology grant, the Sloan Foundation Matter-to-Life grant G-2024-22449 (M.M.L.) and a National Science Foundation Graduate Research Fellowship (E.A.). Some graphical elements were adapted from BioRender.
\end{acknowledgments}

\appendix
\section{Relationship between applied driving forces and chemical potential differences}
\label{sec:theta_mu_rel}

Consider a loop in a state-space network where edges are indexed by $ ij $ (indicating an edge between vertex $ i $ and vertex $ j $). The sum of the chemical potential differences along each edge in the loop is given by:
\begin{align}
    & \sum_{ij \in \text{ loop}} \Delta \mu_{ij} = \sum_{ij} k T \ln \left( \dfrac{J_{ij}}{J_{ji}} \right) = k T \sum_{ij} \ln \left( \dfrac{p_i \:  \rateconstsymbol_{ij} }{p_j \: \rateconstsymbol_{ji} } \right) \\
    &= k T \left[ \sum_{ij} \ln \left( \dfrac{\rateconstsymbol_{ij}/\kappa_{ij}}{\rateconstsymbol_{ji}/\kappa_{ji}} \right) + \ln \left( \prod_{ij} \dfrac{p_i}{p_j} \right) + \ln \left( \prod_{ij} \dfrac{\kappa_{ij}}{\kappa_{ji}} \right) \right] \nonumber
\end{align}
We note that for a sum over the edges in a loop, the probability of each state enters once in the numerator and once in the denominator, and that for every pathway connecting states $ m $ and $ n $, the ratios of the forward and reverse product of equilibrium rate constants is unity:
\begin{align*}
    \dfrac{\kappa_{m,m+1} \kappa_{m+1,m+2} \cdots \kappa_{n-2,n-1} \kappa_{n-1,n}}{\kappa_{m+1,m} \kappa_{m+2,m+1} \cdots \kappa_{n-1,n-2} \kappa_{n,n-1}} &= 1
\end{align*}
Therefore,
\begin{align}
    \sum_{ij \in \text{ loop}} \Delta \mu_{ij} &= k T \sum_{ij \in \text{ loop}} \left( \theta_{ij} - \theta_{ji} \right)
\end{align}

\section{A circuit representation enabling efficient numerical experimentation}
\label{sec:circuit}

For systems of nontrivial complexity, obtaining general insights into steady-state behavior using the rate constant framework is typically analytically intractable \cite{schnakenberg1976network}.
It is convenient to transform variables from rate constants to an equivalent \textit{circuit representation} described by the parameters of the equilibrium system and the thermodynamic forces that drive the system away from equilibrium \cite{lin2020circuit}.

In this representation, each state $ i $ has an associated free energy $ G_i $; the free energies of states $ i $ and $ j $ are related to equilibrium rate constants by:
\begin{align}
    \kappa_{ij} e^{-G_i/k T} = \kappa_{ji} e^{-G_j/k T},
\end{align}
where $ kT $ is the product of the Boltzmann constant and the temperature, which sets the energy scale for the system. The free energies provide a complete description of systems at equilibrium because the probability of such a system visiting any state $ i $ relative to any other state $ j $ is given by the Boltzmann distribution: $ p_i = p_j e^{(G_j-G_i)/k T}$. The currents that emerge in response to non-equilibrium driving are tuned by the resistance $ R $ associated with each transition, defined by:
\begin{align}
    R_{ij} = e^{G_i / k T} / \kappa_{ij} = R_{ji}
\end{align}
Intuitively, the resistance between two states is inversely proportional to the forward flux (equal to the reverse flux) in the equilibrium system. Note that the forward flux from state $ i $ to $ j $ at equilibrium (equal to the reverse flux from $ j $ to $ i $) is equal to $ Z/R_{ij} $ where the normalization constant $ Z $ is the partition function. The driving force $ \theta_{ij} $ in non-equilibrium systems is given by $\theta_{ij} = k T \ln \left( r_{ij} / \eqrateconstsymbol_{ij} \right) $ (Main text Eq. 2), and this takes the form of a battery in the circuit representation \cite{lin2020circuit}.

\begin{figure*}[!htpb]
    \centering
    \includegraphics[width=0.8\textwidth,center]{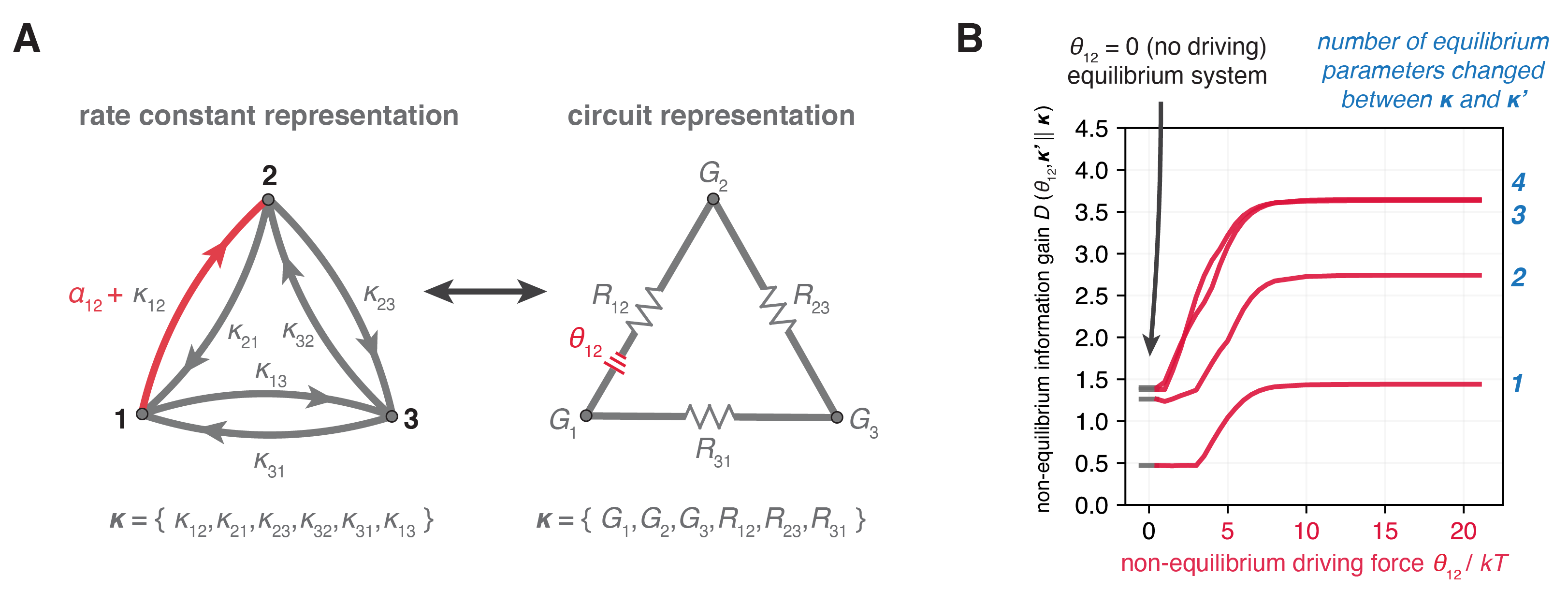}
    \caption{\small \textbf{Equilibrium versus non-equilibrium computation.} (A) The rate constant and circuit representations of a Markov model are equivalent to one another. The transition, rate constant, and voltage shown in red are equivalent representations of the single non-equilibrium element in this system. The vector $ \bk $ contains all the passive parameters for this three-state system. (B) Maximum information gain between systems differing by 1-4 passive parameters as a function of the applied non-equilibrium driving force. The maximum was calculated over a sample of 10 000 sets of parameters $\bk$ and their respective modified versions $\bk'$.}
    \label{fig:figS1}
\end{figure*}

The rate constant representation and the circuit representation encode equivalent information, as illustrated in a simple three-state model (Fig. \ref{fig:figS1}A). In the rate constant representation, the set of all passive parameters in the system, $ \bk $, consists of the passive rate constants $ \kappa_{ij} $ for all $ ij $. In the circuit representation, $ \bk $ consists of the free energy $G_i$ of each state $ i $ and the resistance $ R_{ij} $ between each pair of adjacent states $ i $ and $ j $. The circuit representation naturally decomposes system parameters into independently tunable categories: passive parameters which control equilibrium behavior (the free energies), passive parameters which tune probability flows when the system is not at equilibrium (the resistances), and active parameters (applied driving forces). This parameterization ensures that all free energies and resistances can be independently tuned while preserving detailed balance in the absence of applied driving forces; this enables efficient numerical sampling of different reference equilibrium systems.

We used this parameterization to illustrate the difference between equilibrium versus non-equilibrium computation in the simple three-state model shown in Fig. \ref{fig:figS1}A. We randomly generated a large number of sets of passive parameters $ \bk $, performed a computational step by changing 1-4 individual parameters to yield a new set $ \bk' $, applied a fixed non-equilibrium driving force $ \theta_{12} $ between states 1 and 2 in each network, and calculated the information gain as a function of the number of parameters changed and $ \theta_{12} $. We then examined the maximum information gain, over the different sets of $ \bk, \bk' $ pairs, as a function of $ \theta_{12} $ (Fig. \ref{fig:figS1}B). At equilibrium ($\theta_{12} = 0$), information gain increases modestly as more passive parameters are allowed to change. Regardless of how many parameters are changed, the presence of an applied driving force ($\theta_{12} > 0$) enhances the maximum information gain compared to the same computational step at equilibrium ($\theta_{12} = 0$). The information gain is sensitive to the applied driving force within a limited range, beyond which it saturates. These data suggest that non-equilibrium computation may be needed to achieve a target level of information gain if the number or dynamic range of passive parameters that can be modulated is limited -- a plausible constraint in real biological systems.

\section{Proof of the probability amplification/suppression limit}
\label{sec:amp_lim_proof}

Let us consider an arbitrary network (or graph) $ G $ with $ N $ states. Following \cite{schnakenberg1976network}, we define a maximal tree $ T(G) $ of our network $ G $ to have the following properties:
\begin{enumerate}
    \item $ T(G) $ is a covering subgraph of $ G $, that is, all edges of $ T(G) $ are edges of $ G $, and $ T(G) $ contains all vertices of $ G $;
    \item $ T(G) $ is connected;
    \item $ T(G) $ is acyclic.
\end{enumerate}
The $ M $ maximal trees are indexed by $ \mu $: $ T^{\mu}(G) $, $ \mu = \left\lbrace 1 .. M \right\rbrace $.
Using each $ T^{\mu}(G) $ we now construct a series of directed spanning trees $ T^{\mu}_i(G) $, $ \left\lbrace 1 .. N \right\rbrace $ by assigning a direction to each of its edges so that they all point towards vertex $ i $. Therefore, each edge is associated with a directed rate constant pointing in the direction of the edge.

We now define a function $ \Pi (T^{\mu}_i(G)) $, which gives the product of all the associated directed rate constants in $ T^{\mu}_i(G) $, which we call the \textit{tree product}.
From Kirchoff's matrix-tree theorem we have that the steady-state probability of state $ i $, which we denote by $ p_i $, is given by
\begin{align}
    p_i &= \dfrac{S_i}{S}, \: \text{ where } S_i \equiv \sum_{\mu = 1}^{M} \Pi (T_i^{\mu}(G)) \text{ and } S \equiv \sum_{i = 1}^{N} S_i
\end{align}
This is true whether or not the system is at equilibrium. Define $A_i^{\mu}$ to be the reference equilibrium tree product of tree $\mu$ rooted at $i$, and $A_i^{\mu} \Gamma_i^{\mu}$ to be the corresponding non-equilibrium tree product. (i.e. $ \Gamma_i^{\mu} $ is the ratio of the non-equilibrium tree product to the equilibrium tree product.) Then, the non-equilibrium probability of state $ i $ can be expressed as: 
\begin{equation}
    p_i = \dfrac{\sum_{\mu}A_i^{\mu}\Gamma_i^{\mu}}{\sum_{\mu}\sum_{j}A_j^{\mu}\Gamma_j^{\mu}}
    \label{eq:amplimitA}
\end{equation}
For a fixed spanning tree $\mu$, changing the vertex from $i$ to $j$ can only change the tree product in the terms along the path within the spanning tree that connects $i$ and $j$. Therefore, 
\begin{equation}
    \dfrac{\Gamma_j^{\mu}}{\Gamma_i^{\mu}} = \dfrac{ \prod_{mn \in (i \to j \mid \mu)} \left( \frac{\rateconstsymbol_{mn}}{\eqrateconstsymbol_{mn}} \right) }{ \prod_{nm \in (j \to i \mid \mu)} \left( \frac{\rateconstsymbol_{nm}}{\eqrateconstsymbol_{nm}} \right) } \nonumber
    \label{eq:amplimitB}
\end{equation}
where the product over $ mn \in (i \to j \mid \mu) $ indicates the edges (indexed by $ mn $) present in the path within spanning tree $ \mu $ that connects vertex $ i $ to vertex $ j $. Thus,
\begin{equation}
    e^{-\frac{\Theta}{k T}} \leq \dfrac{\Gamma_j^{\mu}}{\Gamma_i^{\mu}} \leq e^{\frac{\Theta}{k T}}
    \label{eq:amplimitC}
\end{equation}
$ \Theta $ is the maximum sum of $ \ln \left( \frac{\rateconstsymbol_{mn}}{\eqrateconstsymbol_{mn}} \right) - \ln \left( \frac{\rateconstsymbol_{nm}}{\eqrateconstsymbol_{nm}} \right) $ along any path across all spanning trees (Eq. \ref{eq:Theta_definition} in the main text).
Therefore,
\begin{equation}
    p_i \leq e^{\frac{\Theta}{k T}}\dfrac{\sum_{\mu}A_i^{\mu}\Gamma_i^{\mu}}{\sum_{\mu}\sum_{j}A_j^{\mu}\Gamma_i^{\mu}} \nonumber
    \label{eq:amplimitD}
\end{equation}	
Note that
\begin{equation}
    \dfrac{A_i^{\mu}}{A_j^{\mu}} = e^{\Delta G_{ij}/k T}, \nonumber
    \label{eq:amplimitE}
\end{equation}	
where $\Delta G_{ij}$ is the equilibrium free energy difference between states $i$ and $j$, and is thus independent of $\mu$. Therefore,
\begin{equation}
    p_i \leq e^{\frac{\Theta}{k T}}\dfrac{\sum_{\mu}A_i^{\mu}\Gamma_i^{\mu}}{\sum_{\mu}\sum_{j}e^{-\Delta G_{ij}/k T}A_i^{\mu}\Gamma_i^{\mu}} \nonumber
    \label{eq:amplimitF}
\end{equation}
Furthermore, noting that $\sum_{j}e^{-\Delta G_{ij}/k T} = 1/p_i^*$, where $p_i^*$ is the equilibrium probability of state $i$, we finally have:
\begin{equation}
    p_i \leq p_i^* e^{\frac{\Theta}{k T}}\dfrac{\sum_{\mu}A_i^{\mu}\Gamma_i^{\mu}}{\sum_{\mu}A_i^{\mu}\Gamma_i^{\mu}} = p_i^* e^{\frac{\Theta}{k T}}. \nonumber
    \label{eq:amplimitG}
\end{equation}
Following the same logic in reverse, we have that
\begin{align}
    p_i &\geq e^{-\frac{\Theta}{k T}}\dfrac{\sum_{\mu}A_i^{\mu}\Gamma_i^{\mu}}{\sum_{\mu}\sum_{j}A_j^{\mu}\Gamma_i^{\mu}} \nonumber \\
    & \quad = e^{-\frac{\Theta}{k T}}\dfrac{\sum_{\mu}A_i^{\mu}\Gamma_i^{\mu}}{\sum_{\mu}\sum_{j}e^{-\Delta G_{ij}/k T}A_i^{\mu}\Gamma_i^{\mu}} = p_i^* e^{-\frac{\Theta}{k T}} \nonumber
\end{align}
Thus, we have upper and lower bounds for the change in the probability of a single state $ i $ upon driving:
\begin{align}
    e^{-\frac{\Theta}{k T}} \leq \dfrac{p_i}{p_i^*} \leq e^{\frac{\Theta}{k T}}
    \label{eq:amplimit_SI}
\end{align}
This bound is fully general and applies to arbitrary systems. It considers the probability of a given state in isolation, and therefore is purely multiplicative. However, there is an additional constraint on probabilities both in the presence and absence of driving: the probabilities of all states must sum to unity. This can, for specific choices of rate constants, further restrict probability amplification and suppression.

\section{Alternate derivation of probability amplification-suppression bound}
\label{sec:amp_lim_connection}

From Kirchoff's matrix-tree theorem we have that the steady-state probability of state $ i $, which we denote by $ p_i $, is given by
\begin{align}
    p_i &= \dfrac{S_i}{S}, \qquad \text{ where } S_i \equiv \sum_{\mu = 1}^{M} \Pi (T_i^{\mu}(G)) \text{ and } S \equiv \sum_{i = 1}^{N} S_i
\end{align}

This is true whether or not the system is at equilibrium. Define $A_i^{\mu}$ to be the reference equilibrium tree product of tree $\mu$ rooted at $i$, and $A_i^{\mu} \Gamma_i^{\mu}$ to be the corresponding non-equilibrium tree product. (i.e. $ \Gamma_i^{\mu} $ is the ratio of the non-equilibrium tree product to the equilibrium tree product.) Then, the non-equilibrium probability of state $ i $ can be expressed as: 
\begin{equation}
    p_i = \dfrac{\sum_{\mu}A_i^{\mu}\Gamma_i^{\mu}}{\sum_{\mu}\sum_{j}A_j^{\mu}\Gamma_j^{\mu}}
    \label{eq:amplimitH}
\end{equation}

The ratio of any two probabilities is:
    \begin{equation}
    p_j/pi = \dfrac{\sum_{\mu}A_j^{\mu}\Gamma_j^{\mu}}{\sum_{\mu}A_i^{\mu}\Gamma_i^{\mu}}
    \label{eq:amplimitI}
\end{equation}

For a fixed spanning tree $\mu$, changing the vertex from $i$ to $j$ can only change the tree product in the terms along the path within the spanning tree that connects $i$ and $j$. Therefore, 
\begin{equation}
    \dfrac{\Gamma_j^{\mu}}{\Gamma_i^{\mu}} = \dfrac{ \prod_{mn \in (i \to j \mid \mu)} \left( \frac{\rateconstsymbol_{mn}}{\eqrateconstsymbol_{mn}} \right) }{ \prod_{nm \in (j \to i \mid \mu)} \left( \frac{\rateconstsymbol_{nm}}{\eqrateconstsymbol_{nm}} \right) } \nonumber
    \label{eq:amplimitJ}
\end{equation}
where the product over $ mn \in (i \to j \mid \mu) $ indicates the edges (indexed by $ mn $) present in the path within spanning tree $ \mu $ that connects vertex $ i $ to vertex $ j $. Thus,
\begin{equation}
    e^{-\frac{\Theta}{k T}} \leq \dfrac{\Gamma_j^{\mu}}{\Gamma_i^{\mu}} \leq e^{\frac{\Theta}{k T}}
    \label{eq:amplimitK}
\end{equation} $ \Theta $ is the maximum sum of $ \ln \left( \frac{\rateconstsymbol_{mn}}{\eqrateconstsymbol_{mn}} \right) - \ln \left( \frac{\rateconstsymbol_{nm}}{\eqrateconstsymbol_{nm}} \right) $ along any path across all spanning trees (Eq. \ref{eq:Theta_definition}  in the main text).
Therefore,
\begin{equation}
    p_j/p_i \leq e^{\frac{\Theta}{k T}}\dfrac{\sum_{\mu}A_j^{\mu}\Gamma_i^{\mu}}{\sum_{\mu}A_i^{\mu}\Gamma_i^{\mu}} \nonumber
    \label{eq:amplimitL}
\end{equation}	
Note that
\begin{equation}
    \dfrac{A_j^{\mu}}{A_i^{\mu}} = e^{\Delta G_{ji}/k T}=p_j^*/p_i^*, \nonumber
    \label{eq:amplimitM}
\end{equation}	
where $\Delta G_{ji}$ is the equilibrium free energy difference between states $i$ and $j$, and is thus independent of $\mu$. Therefore,
\begin{equation}
    p_j/p_i \leq e^{\frac{\Theta}{k T}} \dfrac{\sum_{\mu}\Gamma_i^{\mu}}{\sum_{\mu}\Gamma_i^{\mu}}p_j^*/p_i^* = e^{\frac{\Theta}{k T}}p_j^*/p_i^*\nonumber
    \label{eq:amplimitN}
\end{equation}
We finally have:
\begin{equation}
    \ln(p_j/p_j^*)-\ln(p_i/p_i^*) \leq \frac{\Theta}{k T}. \nonumber
    \label{eq:amplimitO}
\end{equation}
Because this applies to any pair of states $i$ and $j$, without loss of generality consider $j$ to be an amplified state. Then, there must exist some $i$ whose probability is suppressed due to probability conservation. This implies the weaker bound:
\begin{equation}
    \ln(p_j/p_j^*)\leq \frac{\Theta}{k T}. \nonumber
    \label{eq:amplimitP}
\end{equation}

Swapping $i$ and $j$, we also have the bound on probability suppression:
\begin{align}
    -\frac{\Theta}{k T} \leq \ln(p_j/p_j^*)-\ln(p_i/p_i^*) \leq \frac{\Theta}{k T}
    \label{eq:amplimit_Q}
\end{align}
Note that, in the special case of a single driven transition ($\Theta=\theta$), this coincides with the bound in \cite{cetiner2023universal}.
Swapping $i$ and $j$ on the weaker bound, we obtain the more simplified result highlighted in the main text:
\begin{align}
    -\frac{\Theta}{k T} \leq \ln(p_i/p_i^*) \leq \frac{\Theta}{k T}
    \label{eq:amplimit_R}
\end{align}

\section{Proof of the information gain limit}
\label{sec:infogain_limit}

Starting from the definition of the non-equilibrium information gain, we use the probability amplification limit and add and subtract identical terms:
\begin{widetext}
\begin{align*}
    \begin{split}
        \infdiv{\Theta;\bk'}{\bk}
        &= \sum_{i=1}^N p_i(\Theta,\bk')\ln{\big[{p_i(\Theta,\bk')}\big]}
        -\sum_{i=1}^N p_i(\Theta,\bk')\ln{\big[{p_i(\Theta,\bk)}\big]} -\sum_{i=1}^N p_i(\Theta,\bk')\ln{\big[{p_i(0,\bk')}\big]} \\
        & \qquad +\sum_{i=1}^N p_i(\Theta,\bk')\ln{\big[{p_i(0,\bk')}\big]} +\sum_{i=1}^N p_i(\Theta,\bk')\ln{\big[{p_i(0,\bk)}\big]}
        -\sum_{i=1}^N p_i(\Theta,\bk')\ln{\big[{p_i(0,\bk)}\big]} \\
        &= \underbrace{\sum_{i=1}^N p_i(\Theta,\bk')\ln{\Big[\dfrac{p_i(\Theta,\bk')}{p_i(0,\bk')}\Big]}}_{< \Theta / k T} -
        \underbrace{\sum_{i=1}^N p_i(\Theta,\bk')\ln{\Big[\dfrac{p_i(\Theta,\bk)}{p_i(0,\bk)}\Big]}}_{> -\Theta / k T}
        + \sum_{i=1}^N p_i(\Theta,\bk')\ln{\Big[\dfrac{p_i(0,\bk')}{p_i(0,\bk)}\Big]}\\
        & \leq \dfrac{2 \Theta}{k T} + \sum_{i=1}^N p_i(\Theta,\bk')\ln{\Big[\dfrac{p_i(0,\bk')}{p_i(0,\bk)}\Big]} \leq \dfrac{2 \Theta}{k T} + \sum_{i=1}^N e^{\frac{\Theta}{k T}} p_i(0,\bk')\ln{\Big[\dfrac{p_i(0,\bk')}{p_i(0,\bk)}\Big]}
    \end{split}    
\end{align*}
\end{widetext}
The maximum information gain per computational step of the machine therefore scales exponentially with the total applied driving force $ \Theta $:
\begin{equation}
    \infdiv{\Theta;\bk'}{\bk} \leq \dfrac{2\Theta}{k T} + \infdiv{0;\bk'}{\bk} e^{\frac{\Theta}{k T}}
\end{equation}
Similarly, we find a lower bound for the non-equilibrium information gain (equivalent to an upper bound for the equilibrium information gain in terms of the non-equilibrium information gain):
\begin{widetext}
    \begin{align*}
        \begin{split}
            \infdiv{0;\bk'}{\bk}
            &= \underbrace{\sum_{i=1}^N p_i(0,\bk')\ln{\Big[\dfrac{p_i(0,\bk')}{p_i(\Theta,\bk')}}\Big]}_{\leq \Theta / k T} -
            \underbrace{\sum_{i=1}^N p_i(\Theta,\bk')\ln{\Big[\dfrac{p_i(0,\bk)}{p_i(\Theta,\bk)}\Big]}}_{\geq -\Theta / k T}
            + \sum_{i=1}^N p_i(0,\bk')\ln{\Big[\dfrac{p_i(\Theta,\bk')}{p_i(\Theta,\bk)}\Big]}\\
            & \leq \dfrac{2 \Theta}{k T} + \sum_{i=1}^N p_i(0,\bk')\ln{\Big[\dfrac{p_i(\Theta,\bk')}{p_i(\Theta,\bk)}\Big]} \leq \dfrac{2 \Theta}{k T} + \sum_{i=1}^N e^{-\frac{\Theta}{k T}} p_i(\Theta,\bk')\ln{\Big[\dfrac{p_i(\Theta,\bk')}{p_i(\Theta,\bk)}\Big]}
        \end{split}    
    \end{align*}
\end{widetext}
Hence, the minimum information gain per computational step of the machine similarly scales exponentially with the total applied driving force $ \Theta $:
\begin{equation}
    \infdiv{0;\bk'}{\bk} \leq \dfrac{2\Theta}{k T} + \infdiv{\Theta;\bk'}{\bk} e^{\frac{\Theta}{k T}}
\end{equation}

Note that in this derivation, we have bounded the amplification and suppression of individual probabilities by the bound in Eq. \ref{eq:amplimit_SI}. However, as noted in Sec. \ref{sec:amp_lim_proof}, there is a further constraint that the probabilities of all states must sum to unity. Thus, not all states can be amplified or suppressed to the maximal extent ($ e^{\pm\Theta/kT} $). Despite the fact that we do not make use of this constraint to further tighten the bound, our optimizations approach the information gain bound reasonably closely, but not asymptotically.

\section{Monte Carlo optimization of state-space networks}
\label{sec:mcopt}

The set of states (vertices) in each state-space network (graph) is denoted by $ V $, and the set of edges is denoted by $ E $.
The steady-state probabilities of each state were expressed analytically in terms of the free energy of each state and the resistances associated with transitions between states. 
Here, a ``system'' denotes a particular set of passive parameters (free energies and resistances) for a state-space network. We sought to identify pairs of systems $ \bk $ and  $ \bk' $ that yielded equilibrium and non-equilibrium steady-state probability distributions that differed from one another by a specific amount.

In particular, we minimized a harmonic penalty associated with the distance of a particular system's probability distributions in the $ \DKLT $ vs. $ \DKLO $ plane to a target point $ \left( \DKLOstar, \DKLTstar \right)$:
\begin{align}
    \begin{split}
        L &= \left[ \DKLO - \DKLOstar \right]^2 + \\
        & \qquad \left[ \DKLT - \DKLTstar \right]^2
    \end{split}
\end{align}
Monte Carlo optimization over the vectors of parameters $ \bk = \left\lbrace G_i \right\rbrace_{i \in V} \cup \left\lbrace R_{ij} \right\rbrace_{ij \in E} $ and $ \bk' = \left\lbrace G_i' \right\rbrace_{i \in V} \cup \left\lbrace R_{ij}' \right\rbrace_{ij \in E} $ was performed with an inverse temperature of $ \beta = 4\times 10^4 - 1 \times 10^5 / k T $ for $ 4 \times 10^3 - 4 \times 10^4 $ time steps in Wolfram Mathematica 12.3. Optimization to target points in inaccessible regions beyond the information gain and noise suppression bounds only yielded points that lay within the accessible region.

\section{Analysis of Ras signaling}
\label{sec:ras}

Parameters for the four-state model of Ras are shown in Table \ref{tab:Rasparams}. The rate constants for the Ras cycle were obtained from Refs. \cite{neal1988kinetic,allin2001monitoring,rudack2012ras} and nucleotide concentrations were obtained from \cite{park2016metabolite}. The equilibrium GTP concentration was calculated from the non-equilibrium reaction quotient and the physiological free energy of GTP hydrolysis, or phosphorylation potential, $ \Delta G_\mathrm{P} $. The phosphate association rate constant $ \kappa_{43} $ was set such that the system reached detailed balance at the equilibrium ratio of [GTP]/[GDP]. GTPase-activating proteins (GAP) increase the rate of GTP hydrolysis $\kappa_{32}$ by a factor of $ 10^5 $ \cite{rudack2012ras}. To implement detailed balance, we assume that this lowering of the free energy barrier causes a concomitant increase in $\kappa_{23}$. Guanine nucleotide exchange factors (GEF), such as \textit{SOS}, stabilize the \textit{apo} state relative to GTP and GDP-bound states, leading to a $ 10^4 - 10^5 $-fold increase in the rates of GTP and GDP dissociation \cite{kunzelmann2011fluorescence}.

\begin{table}[!htpb]
    \small
    \caption{Kinetic and thermodynamic parameters for the Ras model}
    \vspace{2mm}
    \centering
    \begin{tabular}{@{}lllll@{}}
        \toprule
        Parameter                         & Value     & Reference \\ \midrule
        $ \kappa_{12} $ (M$^{-1}$ s$^{-1}$)    & $ 1.40 \times 10^{8} $   & \cite{neal1988kinetic}          \\
        $ \kappa_{21} $ (-GEF) (s$^{-1}$)      & $ 1.00 \times 10^{-4} $  & \cite{neal1988kinetic}          \\
        $ \kappa_{21} $ (+GEF) (s$^{-1}$)      & $ 7.78 \times 10^{-1} $  & \cite{neal1988kinetic,kunzelmann2011fluorescence}          \\
        $ \kappa_{23} $ (-GAP) (s$^{-1}$)      & $ 3.40 \times 10^{-4} $  & \cite{neal1988kinetic}          \\
        $ \kappa_{23} $ (+GAP) (s$^{-1}$)      & $ 3.40 \times 10^{1} $   & \cite{neal1988kinetic,rudack2012ras}          \\
        $ \kappa_{32} $ (-GAP) (s$^{-1}$)      & $ 0.16 \times 10^{-4} $  & \cite{neal1988kinetic}          \\
        $ \kappa_{32} $ (+GAP) (s$^{-1}$)      & $ 1.6 $                  & \cite{neal1988kinetic,rudack2012ras}          \\
        $ \kappa_{34} $ (s$^{-1}$)             & $ 0.102 $                & \cite{allin2001monitoring}      \\
        $ \kappa_{43} $ (s$^{-1}$)             & $ 1.98 \times 10^{-6} $  & n/a                             \\
        $ \kappa_{41} $ (-GEF) (s$^{-1}$)      & $ 4.20 \times 10^{-4} $  & \cite{neal1988kinetic}          \\
        $ \kappa_{41} $ (+GEF) (s$^{-1}$)      & $ 3.27 $                 & \cite{neal1988kinetic,kunzelmann2011fluorescence}          \\
        $ \kappa_{14} $ (M$^{-1}$ s$^{-1}$)    & $ 0.51 \times 10^{8} $   & \cite{neal1988kinetic}          \\ \midrule
        $ \Delta G_{\mathrm{P}} $ (kcal/mol)     & $ \approx -12 $ & \cite{desai1997microtubule}    \\
        $ \left[\mathrm{GTP}\right] $ (M) & $ 6.77 \times 10^{-4} $ & \cite{park2016metabolite}       \\
        $ \left[\mathrm{GTP}\right]_\mathrm{eq} $ (M) & $ 2.39 \times 10^{-12} $ & \cite{park2016metabolite,desai1997microtubule}       \\
        $ \left[\mathrm{GDP}\right] $ (M) & $ 3.02 \times 10^{-5} $ & \cite{park2016metabolite}       \\
        $ \left[\mathrm{P}_\mathrm{i}\right] $ (M) & $ 5 \times 10^{-3} $    & \cite{park2016metabolite}\\
        \bottomrule
    \end{tabular}
    \label{tab:Rasparams}
\end{table}

\begin{figure*}[!htpb]
    \centering
    \includegraphics[width=0.7\textwidth,center]{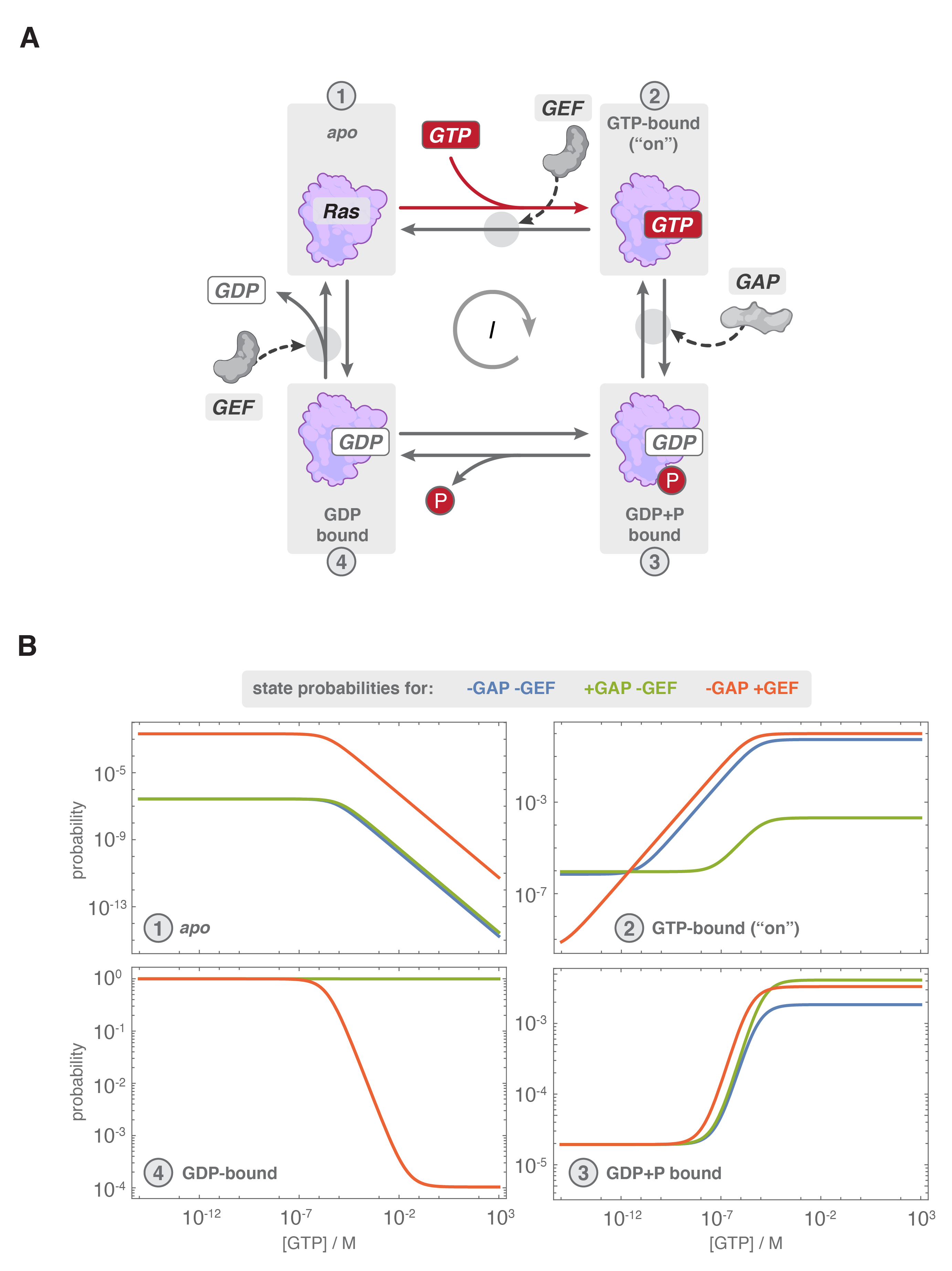}
    \caption{(A) Ras can be modeled as a four-state system: (1) \textit{apo}, with no nucleotide bound; (2) GTP-bound, which is the "on" state, or signaling state; (3) GDP + phosphate bound; and (4) GDP-bound. (B) Probabilities of the individual states are shown as a function of GTP concentration.}
    \label{fig:supp_fig_ras}
\end{figure*}

Using the parameters above, we calculated the probabilities of individual states the presence and absence of GAP and GEF (Fig. \ref{fig:supp_fig_ras}).

We next considered the limits on amplification or suppression of the probability of a single state possible when changing system parameters. Eq. \ref{eq:sigratio_bound} is the most general form of the bound, and it is valid for any arbitrarily driven network. However, we can derive a tighter bound for the specific case of the ``on'' state of Ras:
\begin{align}
    \bigg|\ln{\left(\frac{p_i(\Theta,\bk)}{p_i(\Theta,\bk')}\right)} - \ln{\left(\frac{p_i(0,\bk)}{p_i(0,\bk')}\right)} \bigg| \leq \frac{\Theta}{k T}.
\end{align}

Note that this is because the driven transition leads directly to the ``on'' state, so there does not exist a non-intersecting path in which the driven transition does not point towards the ``on'' state. Thus, its probability cannot be suppressed by driving, i.e. $ 1 \leq p_\mathrm{on} (\Theta,\bk) / p_\mathrm{on} (0, \bk) \leq e^{\Theta / kT} $.

\FloatBarrier
\bibliography{references}

\end{document}